# Magnetic properties of the geometrically frustrated S = 1/2 antiferromagnets, $La_2LiMoO_6$ and $Ba_2YMoO_6$, with the B-site ordered double perovskite structure. Evidence for a Collective Spin Singlet Ground State.

Tomoko Aharen[1], John E. Greedan[1,2], Craig A. Bridges[1], Adam A. Aczel[3], Jose Rodriguez[3], Greg MacDougall[3], Graeme M. Luke[2,3,4], Takashi Imai[2,3,4], Vladimir. K. Michaelis[5], Scott Kroeker[5], Haidong Zhou[6], Chris Wiebe[6,7], Lachlan M.D. Cranswick[8]

[1]Department of Chemistry, McMaster University, Hamilton, Ontario, L8S 4M1, Canada, [2]Brockhouse Institute of Material Research, McMaster University, Hamilton, Ontario, L8S 4M1, Canada, [3]Department of Physics and Astronomy, McMaster University, Hamilton, Ontario, L8S 4M1,Canada, [4]Canadian Institute for Advanced Research, Toronto, Ontario M5G 1Z8, [5]Department of Chemistry, University of Manitoba, Winnipeg, R3T 2N2, Manitoba, Canada, [6]Department of Physics, Florida State University, Tallahassee, 32310-4005, Florida, USA, [7]Department of Chemistry, University of Winnipeg, Winnipeg, MB, R3B 2E9 CANADA, [8]Canadian Neutron Beam Centre, National Research Council, Chalk River Laboratories, Chalk River, Ontario K0J 1J0, Canada,.

**Abstract**

Two B-site ordered double perovskites, $La_2LiMoO_6$ and $Ba_2YMoO_6$, based on the $S = \frac{1}{2}$ ion, $Mo^{5+}$, have been investigated in the context of geometric magnetic frustration. Powder neutron diffraction, heat capacity, susceptibility, muon spin relaxation(μSR), and $^{89}Y$ NMR- including MAS NMR- data have been collected. $La_2LiMoO_6$ crystallizes in $P2_1/n$ with a = 5.59392(19) Å, b = 5.69241(16) Å, c = 7.88029(22), β = 90.2601(30) deg at 299.7K, while $Ba_2YMoO_6$ is cubic, $Fm3m$, with a = 8.39199(65) Å at 297.8 K. $Ba_2YMoO_6$ shows no distortion from cubic symmetry even at 2.8K in apparent violation of the Jahn-Teller theorem for a $t_{2g}^1$ ion. $^{89}Y$ NMR MAS data indicate about a 3% level of Y/Mo site mixing. $La_2LiMoO_6$ deviates strongly from simple Curie-Weiss paramagnetic behavior below 150K and zero-field cooled/ field cooled (ZFC/FC) irreversibility occurs below 20K with a weak, broad susceptibility maximum near 5K in the ZFC data. A Curie-Weiss fit shows a reduced $\mu_{eff}$ = 1.42 $\mu_B$, (spin only = 1.73 $\mu_B$) and a Weiss temperature, $\theta_C$, which depends strongly on the temperature range of the fit. Powder neutron diffraction, heat capacity and $^7Li$ NMR show no evidence for long range magnetic order to 2K. On the other hand oscillations develop below 20K in μSR indicating at least short range magnetic correlations . Susceptibility data for $Ba_2YMoO_6$ also deviate strongly from the C-W law below 150K with a nearly spin only $\mu_{eff}$ = 1.72 $\mu_B$ and $\theta_C$ = - 219(1)K. There is no discernable ZFC/FC irreversibility to 2K. Heat capacity, neutron powder diffraction and μSR data show no evidence for long range order to 2K but a very broad, weak maximum appears in the heat capacity. The $^{89}Y$ NMR paramagnetic Knight shift shows a remarkable local spin susceptibility behavior below about 70K with two components from roughly equal sample volumes, one indicating a singlet state and the other a strongly fluctuating paramagnetic state. Further evidence for a singlet state comes from the behavior of the relaxation rate, $1/T_1$. These results are discussed and compared with those from other isostructural S = ½ materials and those based on S = 3/2 and S = 1.

**Introduction**

Geometric magnetic frustration (GMF) generally originates if spins, constrained by an antiferromagnetic nearest neighbor exchange coupling, are situated on lattices with a topology of triangles or tetrahedra. Magnetic properties of GMF compounds have been studied intensively in recent years due to their exotic ground states, such as spin glasses, spin ices and spin liquids[1]. Since the theoretical proposal by Anderson of one possible model for the spin liquid for frustrated antiferromagnets with *S*=1/2, researchers have been inspired to seek experimental evidence for the existence of spin liquid states [2]. In this model the spins form a collective singlet ground state and the dynamics of singlets is normally retained down to low temperature. The pyrochlore compound $Tb_2Ti_2O_7$[3] and the so-called hyper-kagome compound[4], $Na_4Ir_3O_8$[5], have been proposed as spin-liquid compounds or candidates.

B-site ordered double perovskites with chemical formula $A_2BB'O_6$, where magnetic ions reside on the B'-site, present a face-centered cubic symmetry, which is a three dimensional lattice based on edge sharing tetrahedral and is, thus, geometrically frustrated. In previous work, double perovskites with S = 3/2 and S = 1 spins and both cubic and monoclinic lattice symmetries have been studied in detail from the perspective

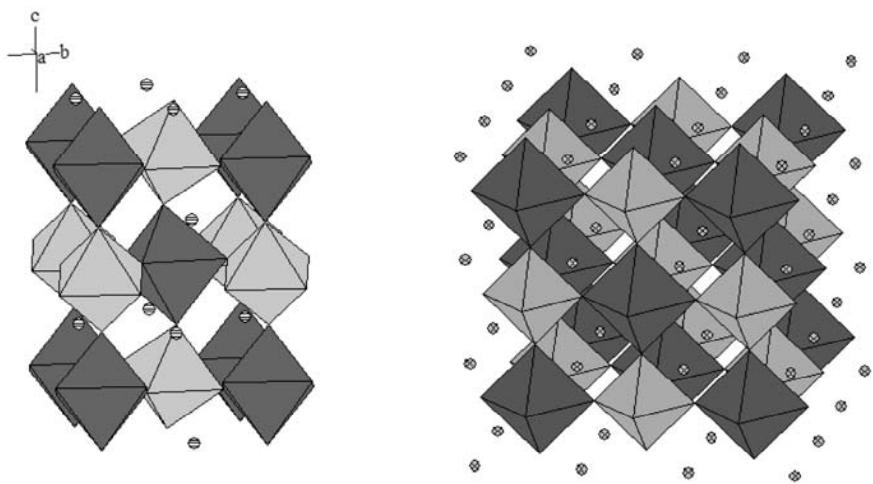

Figure 1. (left) Crystal structure of $La_2LiMoO_6$, $P2_1/n$. Dark grey octahedra represent $MoO_6$ and light grey octahedral, $LiO_6$. Dashed circles represent the La ions. (right) Crystal structure of $Ba_2YMoO_6$, $Fm3m$. Dark grey octahedra represent $MoO_6$ and light grey octahedra, $YO_6$. The small crossed spheres represent the Ba ions.

of geometric frustration. [6,7] Unit cells for the two crystallographic symmetries are shown in Figure 1. For the $S = 3/2$ materials it was found that, while frustration was clearly important, both $Ba_2YRuO_6$ and $La_2LiRuO_6$ did eventually show long range antiferromagnetic order, even in the case of the former where a 1% level of Y/Ru site mixing was detected by $^{89}Y$ MAS (magic angle spinning) NMR. In the case of the $S = 1$ phases, long range order was clearly quenched for both symmetries. $La_2LiReO_6$ showed a very unusual singlet ground state, while $Ba_2YReO_6$, which retains cubic symmetry in apparent violation of the Jahn-Teller theorem, undergoes spin freezing. Paradoxically, Y/Re site mixing was undetectable via $^{89}Y$ MAS NMR. In this work, two $S = ½$ double perovskites, the cubic $Ba_2YMoO_6$ and monoclinic $La_2LiMoO_6$, have been studied in detail using a wide range of probes including powder neutron diffraction, neutron inelastic scattering, bulk magnetic susceptibility, heat capacity, muon spin relaxation and $^{89}Y$ NMR, both MAS and paramagnetic Knight shift measurements. $Ba_2YMoO_6$ had been reported previously to remain paramagnetic to 2K while no magnetic properties studies exist for $La_2LiMoO_6$.[9,10] The results are compared with other isostructural S = ½ B-site ordered double perovskites such as the monoclinic spin glass, $Sr_2CaReO_6$, the cubic ferromagnet, $Ba_2NaOsO_6$ and the cubic antiferromagnet, $Ba_2LiOsO_6$.[11, 12]

**Experimental**

Polycrystalline samples of $La_2LiMoO_6$ and $Ba_2YMoO_6$ were prepared using conventional solid state reaction. For $La_2LiMoO_6$, a stoichiometric mixture of pre-fired $La_2O_3$, $Li_2MoO_4$ and $MoO_2$ were ground and sealed in a Mo crucible, heated to 1100°C in an Ar atmosphere, and kept for 48 hours. For $Ba_2YMoO_6$, $BaCO_3$, pre-fired $Y_2O_3$ and $MoO_3$, were weighed out stoichiometrically and the mixture was fired at 950°C for 12hr., re-ground, then heated to 1300°C in 5% $H_2$/ Ar.

Phase purity was assessed using a focusing Guinier-Hägg camera and PanAlytical X-Pert X-ray diffraction apparatus.

Thermal gravimetric analysis (TGA) was carried out on $Ba_2YMoO_6$ by heating in air to determine the oxidation state of Mo. The weight gain was 1.45% compared to the expected 1.44% indicating the $Mo^{5+}$ state to better than 1%.

D.c. magnetic susceptibility as a function of temperature was measured in 0.01-0.05T applied fields within the temperature range of 2K to 300K and isothermal magnetization data were collected for $La_2LiMoO_6$ in the field range from -5T to 5T at 5K and 20K using Quantum Design MPMS SQUID magnetometer at McMaster University.

The powder neutron diffraction measurements were carried out on the C2 diffractometer at the Canadian Neutron Beam Centre, NRC, Chalk River Laboratories. The measurements were performed at 3.3K and 299.7K with neutron wave lengths, λ

=1.32873 and λ =2.36937 for La$_2$LiMoO$_6$, and λ= 1.33052Å and 2.37192Å at 2.8K and 297.8K for Ba$_2$YMoO$_6$. Additionally, higher resolution diffraction data at 2.8K with λ=1.33052Å was collected for this compound. Refinements were done either with the GSAS or FULLPROF software packages.[13,14]

Heat capacity measurements were carried out using a MagLab calorimeter at McMaster University and an Oxford MPMS apparatus at Florida State University. The lattice match compounds La$_2$LiIrO$_6$ and Ba$_2$YNbO$_6$ were prepared using solid state reactions as well. The synthesis method for La$_2$LiIrO$_6$ is described in ref [8]. For Ba$_2$YNbO$_6$ a stoichiometric mixture of BaCO$_3$, pre-fired Y$_2$O$_3$ and Nb$_2$O$_5$ was heated at 1300°C for 36 hours.

Muon spin relaxation (μSR) measurements were carried out on the M20 beamline at TRIUMF in Vancouver. The zero field muon spin relaxation (ZF- μSR) data were collected at various temperatures between 1.66K and 25K for La$_2$LiMoO$_6$, and at 2K,5K, 10K and 20K for Ba$_2$YMoO$_6$. For La$_2$LiMoO$_6$, longitudinal field μSR (LF-μSR) was also collected at 1.66K for the field of 100, 400, 800, 1200 and 1600G.

$^{89}$Y MAS NMR was carried out at the University of Manitoba using a Varian $^{\text{UNITY}}$*Inova* 600 ($B_0$ = 14.1 T) spectrometer operating at a Larmor frequency of 29.36 MHz. Ba$_2$YMoO$_6$ powder was packed in a ZrO$_2$ MAS rotor with a 22 μL fill volume and spun at 20000 ± 6 Hz in a 3.2 mm double-resonance Varian-Chemagnetics probe. Acquisition was carried out using a 20° tip angle ($ν_{rf}$ = 42 kHz), 270 000 co-added transients, and a recycle delay of 0.8 s. Frequency shifts are referenced with respect to 2M Y(NO$_3$)$_3$ at 0.0 ppm.

$^{89}$Y NMR (nuclear spin $I = 1/2$, nuclear gyromagnetic ratio $^{89}\gamma_n/2\pi = 2.0859$ MHz/Tesla), Knight shift measurements were carried out at McMaster University in a fixed magnetic field of $B = 7.7$ Tesla. The NMR lineshape was obtained by applying the standard FFT technique for the spin echo envelope at ~100 K or higher. However, gradual line broadening was observed and the linewidth exceeded the bandwidth of our R.F. excitations at lower temperature. Accordingly, it was needed to convolute multiple FFT envelopes to obtain the entire lineshape below ~100K down to ~63 K. Below ~63 K, due to the extremely broad lineshape, the lineshape was measured point-by-point by integrating the spin echo intensity. It was confirmed that the point-by-point measurements agree well with the convolution of the multiple FFT envelopes at 63 K. The $^{89}$Y nuclear spin-lattice relaxation rate, $1/T_1$, was measured by applying saturation comb pulses prior to a regular spin-echo sequence, and varying the delay time between them. The single exponential fit of the recovery of nuclear magnetization was good, as expected for a nuclear spin $I = 1/2$ nucleus.

**Results and Discussion.**

**La$_2$LiMoO$_6$**

*Crystal structure*

The crystal structure of La$_2$LiMoO$_6$ was confirmed by both x-ray diffraction and neutron diffraction measurements to be monoclinic with space group P2$_1$/n, consistent with the previous X-ray single crystal diffraction result of Tortelier and Gougeon [10]. The unit cell constants refined from the data collected at 300K are a= 5.5939(2)Å, b=5.6924(2)Å, c=7.8803(2)Å and β=90.260(3)°. The overall refinement results are shown in Fig.2 and Table 1 where both datasets were refined simultaneously using GSAS. The model assumes full ordering of the B-site cations (Li and Mo). This is consistent with the empirical study of B site ion ordering in the double perovskites by Anderson *et al* [15] which shows that for a formal charge difference of 4, full B-site ordering is found, regardless of the difference in radii, Δr. For this compound even Δr is significant as the radii for $Mo^{5+}$ (0.61Å) and $Li^+$ (0.76Å) differ by almost 25%.[18] Moreover, the isostructural compound, $La_2LiRuO_6$, was shown to have no detectable Li/Ru disorder using $^7Li$ MAS NMR[19]. There is no indication from the B-site atomic displacement parameters that this is not correct. The crystal structure of this compound at 3.3K was also refined and the monoclinic structure, $P2_1/n$ was retained.

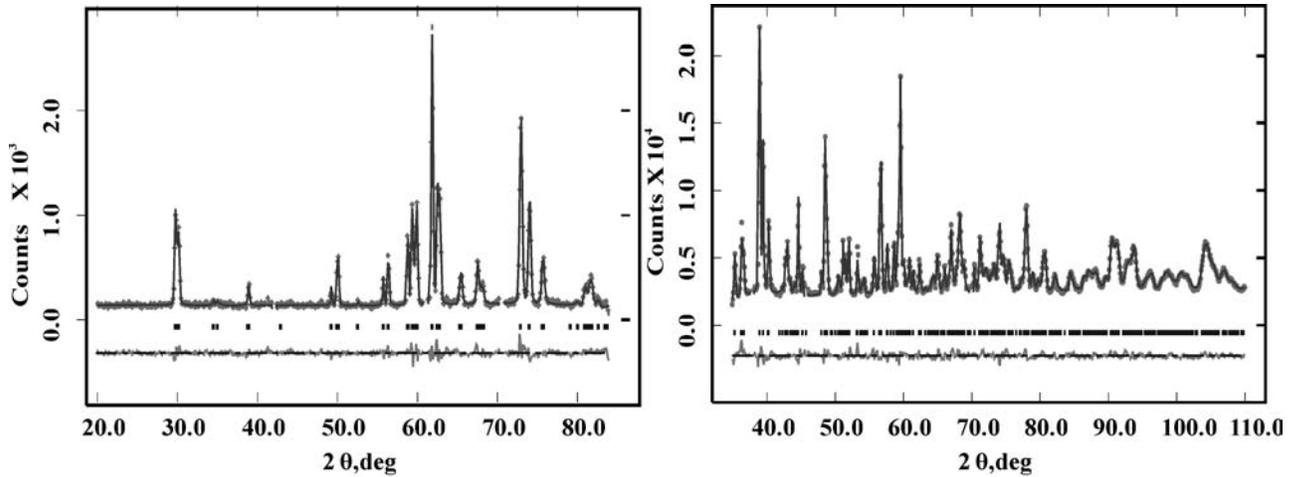

Figure 2. Rietveld refinement of neutron diffraction data collected with two wavelengths, 2.37Å (left) and 1.33 Å(right) at 297K for $La_2LiMoO_6$. The vertical tick marks locate the Bragg peak positions and the lower horizontal line represents the difference between the data,(crosses) and the model (solid line). A weak impurity peak near 70 deg was excluded for the 2.37Å pattern.

Note that in spite of the overall monoclinic symmetry, the nearest neighbor environments of both B-site ions are those of a nearly perfect octahedron with a weak tetragonal compression. Especially for the Mo – O polyhedron, the O – Mo – O angles are either 90 or 180 degrees to within a few tenths of a degree. The weak tetragonal compression will split the nominally $t_{2g}^1$ configuration into a lower orbital singlet ($d_{xy}$) and a higher doublet ($d_{xz}, d_{yz}$).

Table 1a. The results for a Rietveld refinement using GSAS[13] of neutron powder data for La$_2$LiMoO$_6$ (299.7K) in P2$_1$/n. a= 5.5939(2) Å, b = 5.6924(2) Å, c = 7.8803(2) Å, β = 90.260(3)°. R$_{wp}$ = 0.0519, $\chi^2$ = 2.47.

| atom | x | y | z | U$_{iso}$(Å$^2$) |
|------|---|---|---|------------------|
| La | -0.0098(6) | 0.05034(35) | 0.2516(4) | 0.0106(5) |
| Li | 0.5 | 0 | 0 | 0.029(5) |
| Mo | 0.5 | 0 | 0.5 | 0.0119(10) |
| O1 | 0.1917(8) | 0.2159(8) | -0.0460(6) | 0.0137(12) |
| O2 | 0.2855(7) | -0.3077(8) | -0.0421(6) | 0.0105(11) |
| O3 | 0.0836(7) | -0.5207(5) | 0.2387(5) | 0.0108(8) |

Table 1b. Selected interatomic distances (Å) and angles (°) for La$_2$LiMoO$_6$ at 299.7K.

| bond | dis. / angle |
|------|--------------|
| Mo-O1(x2) | 1.975(4) |
| Mo-O1(x2) | 1.966(4) |
| Mo-O3(x2) | 1.940(4) |
| Li-O1(x2) | 2.148(4) |
| Li-O2(x2) | 2.148(4) |
| Li-O3(x2) | 2.117(4) |
| O1-Mo-O1 | 180.0 |
| O1-Mo-O2 | 89.02(23) |
| O1-Mo-O3 | 90.03(15) |
| O2-Mo-O2 | 179.96 |
| O2-Mo-O3 | 89.72(18) |
| O3-Mo-O3 | 180.0 |
| Li-O1-Mo | 150.93(24) |
| Li-O2-Mo | 151.86(25) |
| Li-O3-Mo | 152.42(20) |

*Magnetic properties*

The susceptibility of this compound is shown in Fig. 3. Note the deviation from the Curie-Weiss law below ~ 150K. Data were fitted to the Curie-Weiss law, $\chi = C/(T-\theta)$ for the range T >150K, yielding, $\theta$ = -45(2)K and C = 0.253(1) (emu/mol-K) corresponding to $\mu_{eff}$ = 1.42 $\mu_B$, compared to the spin only value of 1.73 $\mu_B$. A ZFC/FC divergence was observed at 20K and a broad peak around 5K. It should be noted that there was some sample to sample variability for this material, especially in terms of the derived $\theta$ value which varied from ~ -10K to -45K. Nonetheless, robust observations for all samples include the deviation from the C-W law below 150K, the ZFC/FC divergence at 20K and the broad maximum near 5K. Thus, an average frustration index, $f = |\theta|/T^*$, falls in the range ~ 1 - 2 taking the divergence temperature as $T^*$, in stark contrast to the isostructural S = 3/2 and S = 1 materials, $La_2LiRuO_6$ and $La_1LiReO_6$, where f ~7 for each.[6, 7] Thus, from the bulk susceptibility there is no evidence for magnetic frustration, at least from the frustration index.

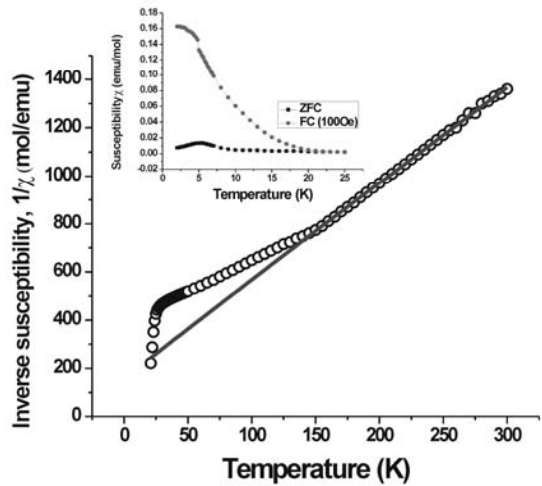

Figure 3. The inverse magnetic susceptibility of $La_2LiMoO_6$ showing Curie-Weiss behavior below 150K. (Inset) The susceptibility below 25K at an applied field of 100Oe. Note the zfc/fc divergence below ~ 20K and the broad maximum at 5K in the zfc data.

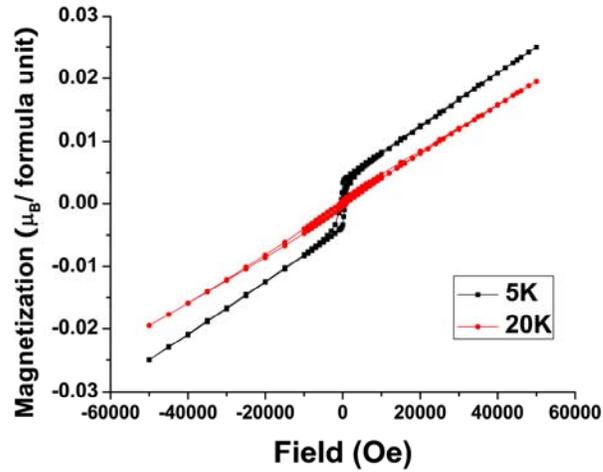

Figure 4. Hysteresis loops for La$_2$LiMoO$_6$ at 5K and 20K.

The field dependence of the susceptibility shown in Figure 4 discloses a weak hysteresis at 5K, which disappears at 20K, indicating the presence of a weak spontaneous moment below 20K.

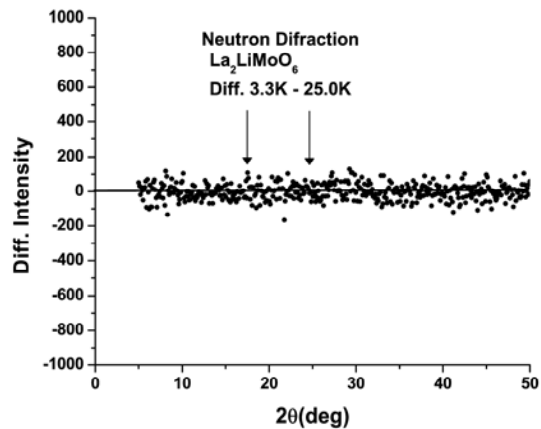

Figure 5. The difference neutron diffraction pattern for La$_2$LiMoO$_6$, 3.3K – 25.0K. The arrows show the positions expected for magnetic reflections assuming a Type 1 fcc magnetic structure as found for La$_2$LiRuO$_6$.[19]

*Magnetic Neutron Diffraction.*

There is no sign of long range order in the neutron powder diffraction data, Fig. 5, where the difference plot, 3.3K – 25K is shown. Of course the intensities of magnetic Bragg peaks for S = ½ spins is expected to be ~ 10 times weaker than those from the S = 3/2 analog, La$_2$LiRuO$_6$, where such peaks were readily visible above background.[19]

*Heat Capacity.*

Heat capacity data, Fig. 6 shows only broad features at 20K and 5K, consistent with the bulk susceptibility and supportive of the absence of long range magnetic order. Unfortunately, the chosen lattice match material, La$_2$LiIrO$_6$ was not useful here as its heat capacity actually exceeded that for LaLiMoO$_6$ above 20K. The entropy removal from 7K – 2K, including the peak at 5K, is 0.25 J/mol-K$^2$ or 4.4% of that expected for S = 1/2.

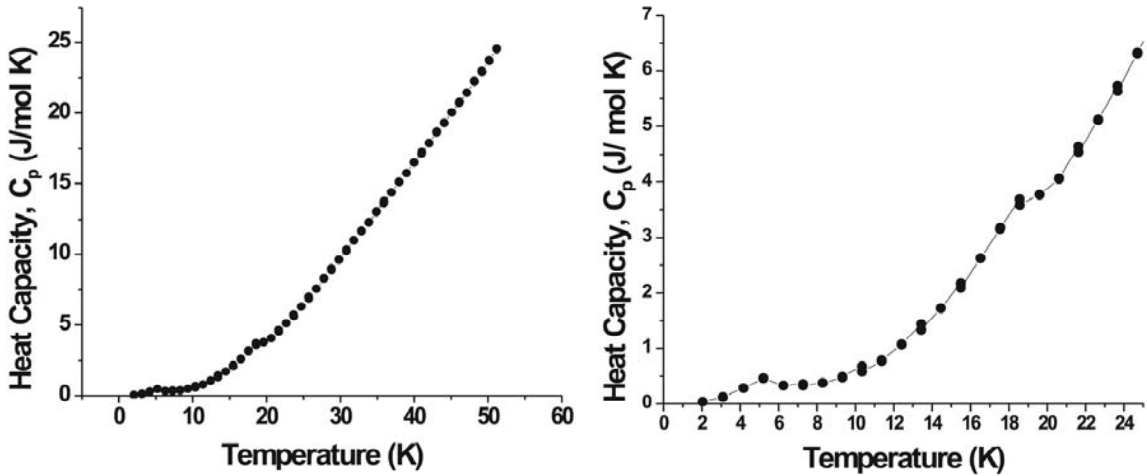

Figure 6. (Left)The heat capacity C$_p$ as a function of temperature for La$_2$LiMoO$_6$. (Right) The region below 25K showing two broad peaks at ~ 18K and 5K.

*Muon Spin Relaxation.*

Muon spin relaxation (μSR) is a local probe of magnetism and is therefore complementary to the bulk techniques just described. Zero-field (ZF) μSR data are shown in Fig. 7(a). Although one observes only weak, paramagnetic relaxation at 25 K, well-defined oscillations are observed at 2 K. The latter indicates that each muon senses a static, unique field in the low temperature regime. A fast Fourier transform (FFT) of the data at 6 K detects three peak frequencies, and this is consistent with the presence of three magnetically-inequivalent oxygen sites in the crystal structure (see Table 1). As a result, the data was fit to the following function:

$A_oP(t) = A_1\exp(-\omega_1 t)\cos(\omega_1 t+\theta) + A_2\exp(-\omega_2 t)\cos(\omega_2 t+\theta) + A_3\exp(-\omega_3 t)\cos(\omega_3 t+\theta) + A_{tail}\exp(-\lambda_{tail} t) + A_G G_{KT}$.

The three frequencies are given by $\omega_1$, $\omega_2$, and $\omega_3$, respectively, and their temperature-dependences are plotted in Fig. 7(b). The onset of these frequencies is around 20 K, in line with the susceptibility and heat capacity data. The $A_{tail}$ term accounts for the volume fraction of the sample that is ordered but with a local field parallel to the muon spin. Finally, the Gaussian Kubo-Toyabe term (temperature-independent relaxation rate was assumed in this analysis) accounts for the volume fraction of the sample that is not ordered. Note that the amplitude of this component quickly dropped to zero below ~ 20 K, and so 100% of the volume fraction was found to be ordered at low temperatures.

For the data collected below 5K, an additional frequency peak was observed in the FFTs at ~ 2 MHz shown in the inset to Fig. 7b. This is likely related to the broad 5K peak in the susceptibility and heat capacity and may signify a change in the local magnetic structure. As well, longitudinal field data (not shown) at 2K show that decoupling occurs for an applied field of 800 Oe, indicating a static spin ground state.

While the behavior seen for $La_2LiMoO_6$ could be taken as evidence for long range magnetic order, the evidence from heat capacity and neutron diffraction indicate otherwise. Of course it is not possible to measure a correlation length for the magnetic order sensed by a local probe such as μSR and it is thus not possible to distinguish between true long range order and order on a finite but significant length scale. A similar situation was encountered with $Li_2Mn_2O_4$ for which strong oscillations were detected by μSR, whereas, the length scale of the magnetic correlations is known to be finite but of the order 90 Å. [20]

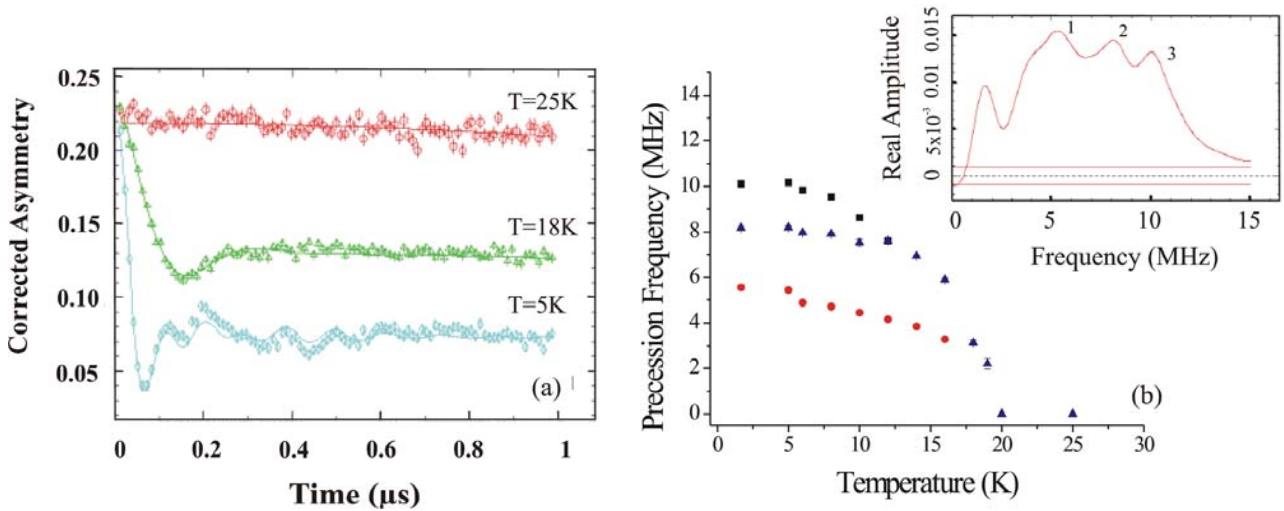

Figure 7. a. (Left) μSR results for $La_2LiMoO_6$. Corrected asymmetry at selected temperatures. b.(Right) Temperature dependence of the amplitudes of the three frequencies (1-3) obtained from a fast fourier transform (see text) of the data. The inset shows the result at 1.7K where a fourth, low amplitude frequency at ~ 2MHz is seen. This feature disappears above 5K. (Color online)

**Ba$_2$YMoO$_6$**

*Crystal Structure*

The neutron diffraction data collected at 2.7K and 298K with λ = 1.33 Å and 2.37 Å were refined and the results are shown in Fig. 8 and Table. 2. The results are in excellent agreement with those of Cussens et al. [9] According to the refinement, the compound retains cubic structure with S.G. Fm3m, and the lattice constant was determined to be 8.3827(7) Å at 2.7K. Surprisingly, even at 2.7K, there is no clear distortion observed this compound although one could expect to see Jahn-Teller distortion arising from $d^1$ electronic configuration of $Mo^{5+}$. Additionally, no distortion was observed with higher resolution data, ($\Delta d/d \sim 2 \times 10^{-3}$) collected at 2.7K, Figure 8(bottom). While the symmetry consequences of a J-T distortion in this material are not clear, certainly a lowering of the m-3m site symmetry at the Mo site must occur. This will likely result in a lowering of the space group symmetry. Such an effect was not observed in our data but a very subtle distortion below the resolution of the data cannot of course be ruled out. It is important to note that room temperature x-ray diffraction data of much better resolution, $\Delta d/d \sim 1 \times 10^{-3}$, also did not show any distortion from Fm3m symmetry.

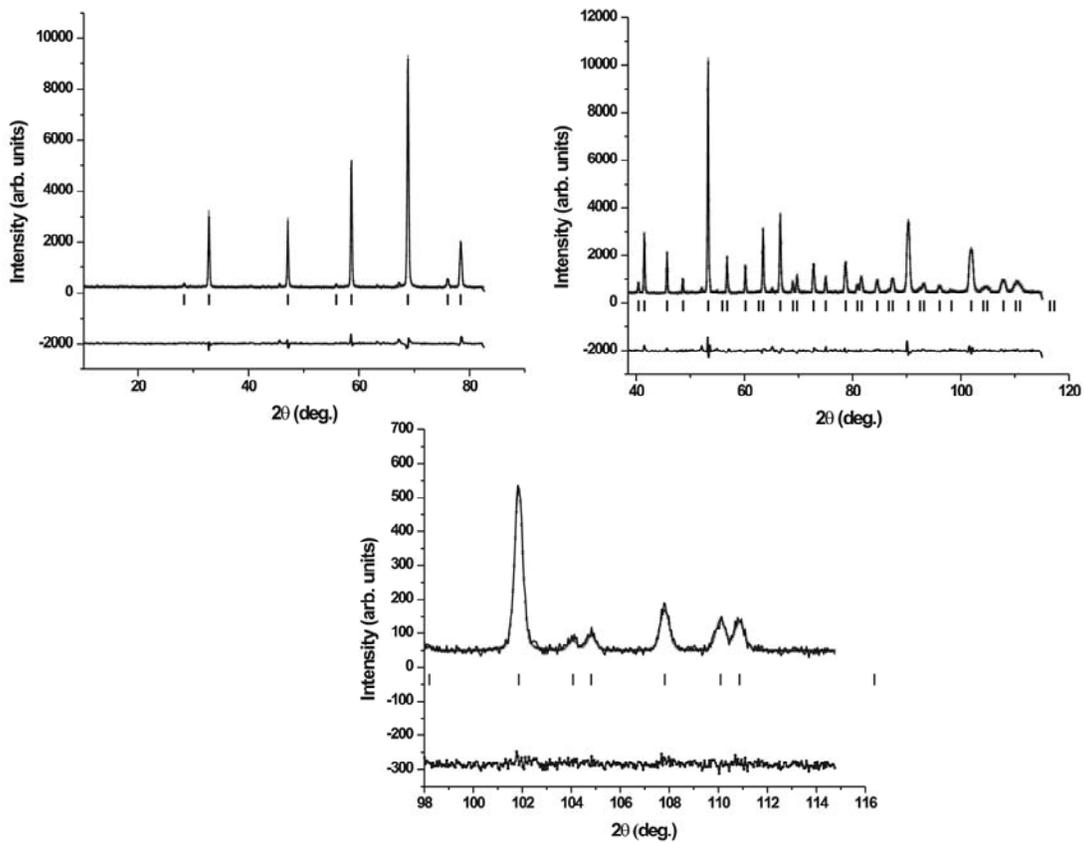

Figure 8. (Top) The refinement result of neutron diffraction pattern for two wavelengths, 2.37Å (left) and 1.33 Å(right), collected at 2.7K for Ba$_2$YMoO$_6$. (Bottom) Higher resolution( Δd/d ~ 2 x 10$^{-3}$), high angle data showing that cubic symmetry is retained. The vertical tick marks locate the Bragg peaks and the lower horizontal line is the difference between the data (circles) and the fit (solid line).

Table 2a. The results for a Rietveld refinement using FULLPROF[14] of neutron powder data for $Ba_2YMoO_6$ in Fm3m at 297.8K and 2.7K. a = 8.3920(6) Å (297.8K), a = 8.3784(6) Å (2.7K). Agreement indices at 297.8K: $R_{wp}$ = 0.063[0.076], $\chi^2$ = 2.51[2.08], $R_B$ = 0.031[0.021] and at 2.7K: $R_{wp}$ = 0.067[0.054], $\chi^2$ = 3.00[9.45], $R_B$ = 0.035[0.011]. Numbers in [ ] are those from refinement of the λ=2.37Å data and the others from λ=1.33Å data.

| 297.8K | x | y | z | $U_{iso}$(Å$^2$) | $U_{11}$ | $U_{22}$ | $U_{33}$ |
|---|---|---|---|---|---|---|---|
| Ba | 0.25 | 0.25 | 0.25 | 0.0072 | 0.0020(2) | 0.0020(2) | 0.0020(2) |
| Y | 0.5 | 0.5 | 0.5 | 0.0043 | 0.0012(4) | 0.0012(4) | 0.0012(4) |
| Mo | 0 | 0 | 0 | 0.0044 | 0.0012(4) | 0.0012(4) | 0.0012(4) |
| O | 0.2347(2) | 0 | 0 | 0.0123 | 0.0017(4) | 0.0043(3) | 0.0043(3) |

| 2.7K | x | y | z | $U_{iso}$(Å$^2$) | $U_{11}$ | $U_{22}$ | $U_{33}$ |
|---|---|---|---|---|---|---|---|
| Ba | 0.25 | 0.25 | 0.25 | 0.0015 | 0.0005(3) | 0.0005(3) | 0.0005(3) |
| Y | 0.5 | 0.5 | 0.5 | 0.0023 | 0.0006(4) | 0.0006(4) | 0.0006(4) |
| Mo | 0 | 0 | 0 | 0.0008 | 0.0002(1) | 0.0002(1) | 0.0002(1) |
| O | 0.2341(2) | 0 | 0 | 0.0069 | 0.0005(4) | 0.0026(3) | 0.0026(3) |

Table 2b. Selected bond lengths and comparison with the sum of the ionic radii.[16]

| bond | 297.8K | 2.7K | $r_B + r_O$ $r_{B'} + r_O$ |
|---|---|---|---|
| Mo-O x6 (Å) | 1.969(3) | 1.963(2) | 1.96 |
| Y-O x6 (Å) | 2.227(3) | 2.226(2) | 2.25 |

This compound has a charge difference of 2 for the B site ions which differ by 0.29Å in ionic radius [13], thus locating it near the B-site order/disorder boundary in the phase diagram of Anderson et al. [15] and the issue of Y/Mo site disorder should be investigated. This discussion follows closely that already presented for $Ba_2YRuO_6$.[6] Woodward et al have studied B-site ordering in a number of double perovskites of the type $A_2BB'O_6$ with the combination $B^{3+}/B'^{5+}$ which is relevant here. [16,17] The extent of B/B' site order was determined quantitatively by a constrained refinement of neutron powder diffraction data and by monitoring the relative widths of the hkl all-odd supercell reflections to the hkl all-even subcell reflections. In all cases of partial B-site order, the hkl all-odd reflections were significantly broader than the hkl all-even reflections. To summarize some of their findings, for $\Delta r > 0.260$ Å, 100% B/B' -site order is always found. For example, among the Fm3m phases studied, $Ba_2YNbO_6$ is judged to be 100% ordered, $\Delta r = 0.260$ Å, while $Ba_2ScNbO_6$ and $Ba_2ScTaO_6$, with $\Delta r = 0.105$ Å, show only about 50% order. Thus, one would expect 100% B-site order in $Ba_2YMoO_6$.

First, the neutron diffraction data were examined. The contrast in neutron scattering lengths between Mo(6.715fm) and Y(7.75fm) is ~ 15%, about twice the contrast for x-ray scattering. There is no indication from the atomic displacement parameters, U, Table 2, for obvious Y/Mo site disorder nor from the derived interatomic Y-O and Mo-O distances, which agree well with those of Cussen et al. and the sum of the ionic radii.[9,15] As well, the FWHM of several hkl-odd/hkl-even pairs were examined and no significant differences were found. For example the odd/even width ratio for the (551,711)/(624) pair is 1.04(8) and that for the (553,731)/(642) pair is 1.01(8). Thus, from the neutron powder diffraction data, there is no detectable Y/Mo site disorder.

Hence, $^{89}$Y MAS NMR data were collected for this compound. Previous studies on $Ba_2YRuO_6$ which also showed no evidence for B-site disorder from neutron data disclosed a ~ 1% Y/Ru mixing.[6] Figure 9 shows the spectrum obtained for $Ba_2YMoO_6$ at 288K and there are two distinct peaks located at -1391 ppm and -1334ppm with an area ratio of 81%/19%, respectively. The uncertainty in these areas is +/- 3%. This is a considerably greater site mixing than found in $Ba_2YRuO_6$ or $Ba_2YReO_6$.[6,7] Following the previous analysis [6] the more negatively shifted peak is assigned to a Y-(O-Mo)$_6$ local environment and the other to Y-(O-Mo)$_5$(O-Y). Taking into account the coordination number of six at the Y-site, an estimate of the mixing level is ~ 3%, too small to be detected in the neutron diffraction data.[6] Note that these results imply that a finite concentration of "defect" clusters of composition [Mo-(O-Mo)$_6$] will exist and this will likely be reflected in both the bulk and especially the local susceptibility.

*Magnetic properties*

$Ba_2YMoO_6$ shows only apparent bulk paramagnetic behavior down to 2K, Figure 10, which is consistent with the result previously reported by Cussen et al. [9]. The result of Curie-Weiss fitting above 150K shows θ = - 219(1) K, and an effective moment of 1.72(1)$\mu_B$ which is essentially the spin only value for S=1/2 (1.73$\mu_B$). These values are somewhat larger than those reported by Cussen et al. The frustration index, $f = |\theta_{cw}|/T_N$ for this compound is thus > 100, indicative of extremely frustrated behavior. In order to verify any ordering/ spin dynamics, we have conducted low temperature neutron diffraction, heat capacity, muon spin relaxation and $^{89}$Y NMR investigations.

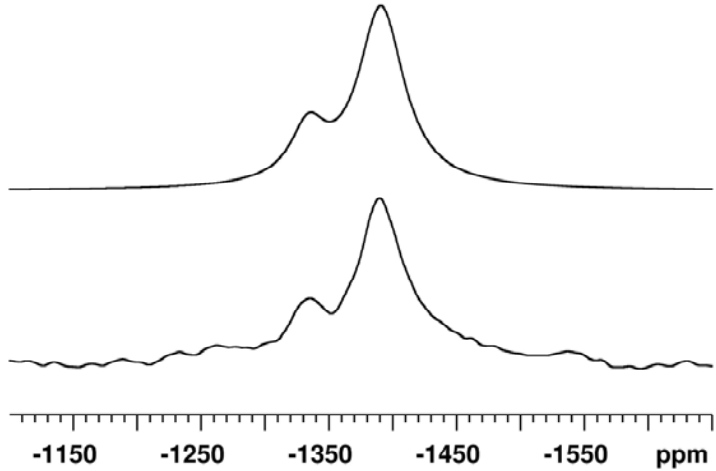

Figure 9. $^{89}$Y MAS(magic angle spinning) NMR of $Ba_2YMoO_6$ at 288K. The top is a simulation and the bottom are the data. The relative intensities of the two peaks is 19%/81% with an error of +/- 3%.

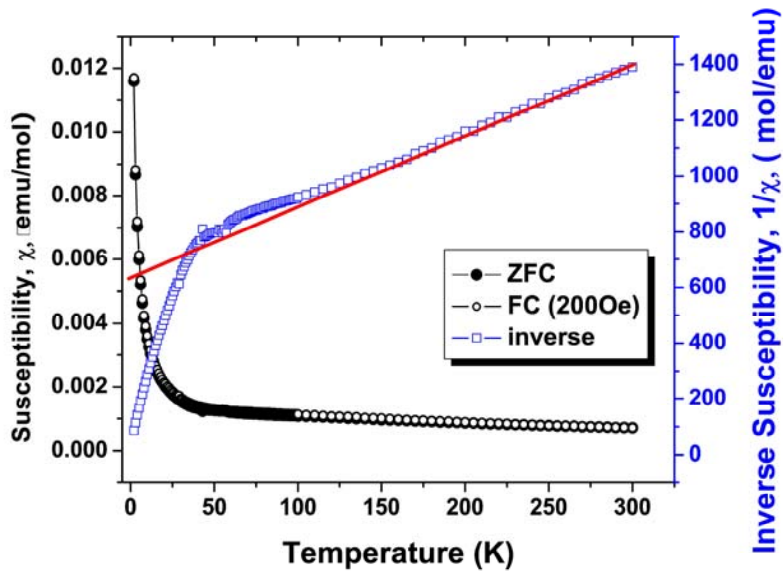

Figure 10. The susceptibility and inverse susceptibility of $Ba_2YMoO_6$ at an applied field of 200 Oe.

*Magnetic Neutron Diffraction.*

From the difference powder neutron diffraction pattern (2.8K – 297.8K), Fig. 11, no magnetic Bragg peaks were detected in the diffraction pattern. As in the case for $La_2LiMoO_6$, while this observation is evidence against long range magnetic order in this system, it is not necessarily conclusive given the small S value and other corroborating data are needed.

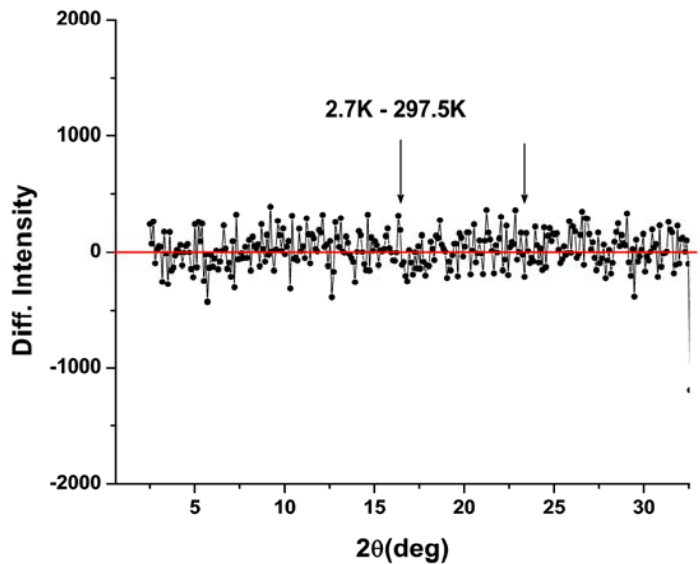

Figure 11. Neutron diffraction difference pattern, 2.7K – 297.5K for $Ba_2YMoO_6$. The arrows show the expected positions of magnetic reflections assuming a Type 1 fcc magnetic structure as found for $Ba_2YRuO_6$.[6]

*Heat Capacity.*

Heat capacity data for Ba$_2$YMoO$_6$ and Ba$_2$YNbO$_6$, the lattice match phase, are shown in Figure 12. Note the absence of a λ-type peak which again is evidence against long range order. The magnetic heat capacity shows very broad peak around 50K. The entropy lost in this temperature range (<50K) was calculated to be 51.9% of the theoretical value (R ln 2).

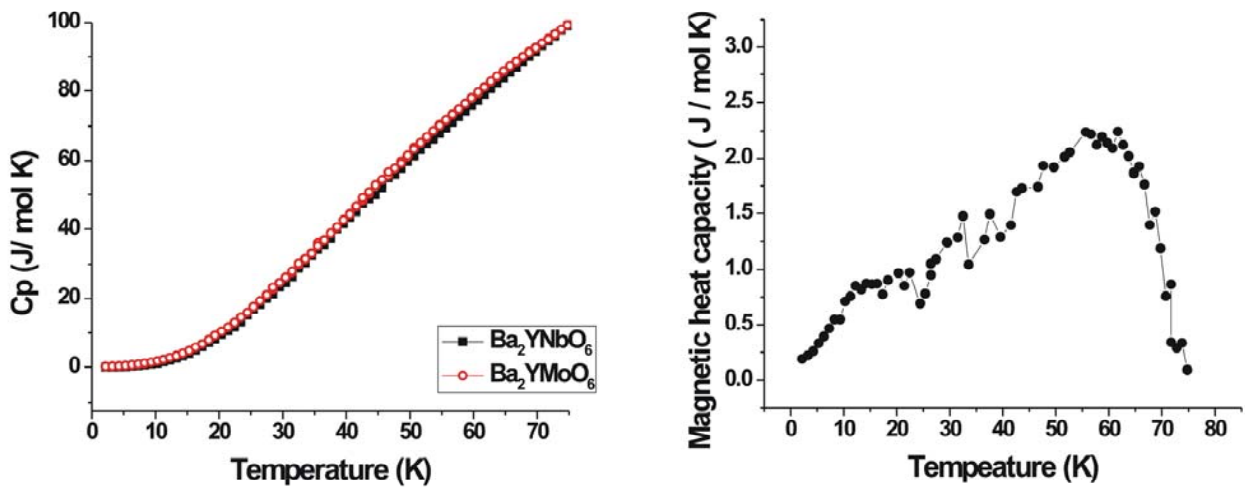

Figure 12. (a.) (Left) The heat capacity of Ba$_2$YMoO$_6$ and Ba$_2$YNbO$_6$.
(b.) (Right) Magnetic heat capacity of Ba$_2$YMoO$_6$.

*μSR.*

μSR data, collected at various temperatures, are shown in Fig. 13. One can see that the relaxation functions indicate dynamic spin behavior down to 2K with a weak slowing down at 2K but no indication of spin freezing or order on any length scale. The relaxation functions were fitted with the equation for dynamic behavior, P(t) = A exp (-λt). Overall, the spins in this compound show persistent fluctuating spin behavior within μSR time window.

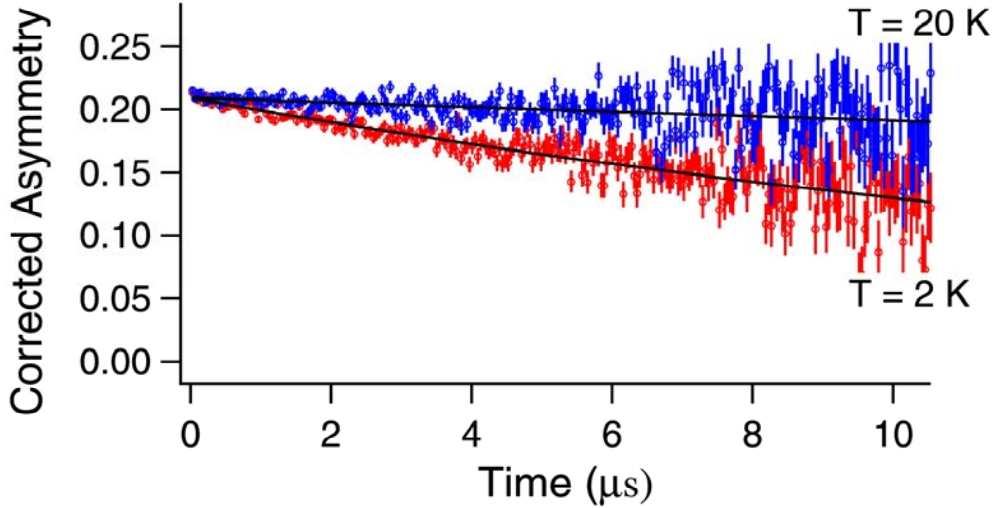

Figure 13. Zero Field (ZF) muon spin relaxation (μ SR) data for $Ba_2YMoO_6$ for two temperatures. The lines are fits to a single exponential relaxation function, see text. (Color online)

*$^{89}$Y NMR.*

$Ba_2YMoO_6$ magnetic properties and spin dynamics were also studied through $^{89}$Y NMR, measuring the linewidth of resonance peaks and paramagnetic Knight shifts. NMR is an extremely useful probe to uncover the distinctive temperature dependences of the local magnetic susceptibility which is different from the overall bulk average. For example earlier $^{35}$Cl and $^{17}$O NMR measurements uncovered the presence of vanishingly small spin susceptibility in a Kagome antiferromagnet $ZuCu_3(OH)_6Cl_2$, although the bulk susceptibility data grows monotonically down to 2 K.[23,24] In Figure 14(a), representative $^{89}$Y NMR lineshapes at 295 K, 185 K, and 75 K are presented. A sharp peak with FWHM (Full Width at Half Maximum) less than 10 kHz is evident near 295 K with however, a broad hump at the higher frequency side, and hence the overall lineshape is tapered toward higher frequency.  In addition, a small side peak on the higher frequency side of the main peak is observed which is most distinctly visible at 185 K near 16.03 MHz.  The main sharp peak broadens gradually with decreasing temperature and masks the presence of this side peak which is no longer distinguishable at 75 K. The entire lineshape is shifted to the lower frequency side compared with the expected position of the resonance in 7.7 Tesla, $^{89}f_o = {^{89}\gamma_n} B$ = 16.061 MHz (shown by grey dotted line).  For example, the peak frequency at 295 K, $^{89}f$ = 16.041 MHz, is shifted from $^{89}f_o$ by $\Delta f = {^{89}f} - {^{89}f_o}$ = -0.020 MHz.  This shift $\Delta f$ is caused by the paramagnetic *Knight shift*, as defined by

$$^{89}K = \Delta f / {^{89}f_o}. \qquad (1)$$

$^{89}K$ is represented in terms of %, by applying a factor of 100, following the common convention in condensed matter physics.  (Note that 0.01% of the Knight shift corresponds to 100 ppm.)  $^{89}K$ is related to the paramagnetic susceptibility of Mo magnetic moments $\chi$ by

$$^{89}K = A \chi + K_{chem}, \qquad (2)$$

where $A$ is the hyperfine coupling constant, and $K_{chem}$ is a small chemical shift (typically, $|K_{chem}| <$ 0.02 %, or equivalently, < 200 ppm).

Upon cooling, the overall lineshape shifts further to lower frequency. In view of the observed increase of the bulk susceptibility data χ at lower temperatures presented earlier in Figure 10, we can understand the observed temperature dependent shift of the NMR lineshapes as the consequence of a negative hyperfine coupling with magnetic moments, *i.e. A < 0*. The observed linewidth at 75 K, FWHM ~ 9 kHz, is typical for a paramagnetic insulator of a large bulk magnetic susceptibility. However, below ~75 K, qualitatively different NMR lineshapes emerge, as shown in Fig.14(b) in the form of two distinct peaks, clearly seen at 50K, with one shifted to higher frequency. At 50 K the main peak is narrower than the higher frequency component only by a factor of ~2 and the

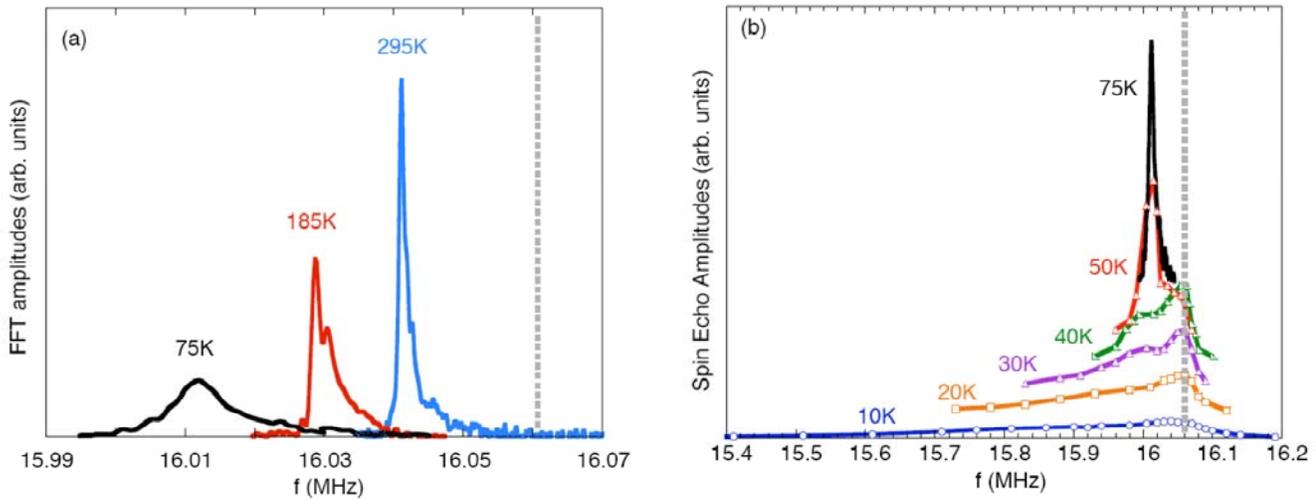

Figure 14. $^{89}$Y NMR lineshape at various temperatures. (a.) (Left) Representative line shapes at selected temperatures for $Ba_2YMoO_6$. (b.) (Right) Evolution of the line shape below 75K. (Color online)

Below 50K the low frequency component becomes very broad with decreasing temperature while shifting dramatically to lower frequencies. The proximity between the two peaks and the asymmetric lineshape makes it difficult to estimate the intensity ratio accurately. Recalling that the hyperfine coupling constant $A < 0$, the $^{89}$Y nuclear spins represented by the low frequency peak sense a growing hyperfine field with decreasing temperature. In contrast the broader peak at the higher frequency shifts to *higher* frequency with decreasing temperature reaching a constant value at ~ 40K. Thus, the local magnetic susceptibility near these $^{89}$Y nuclear spins *decreases* with decreasing temperature.

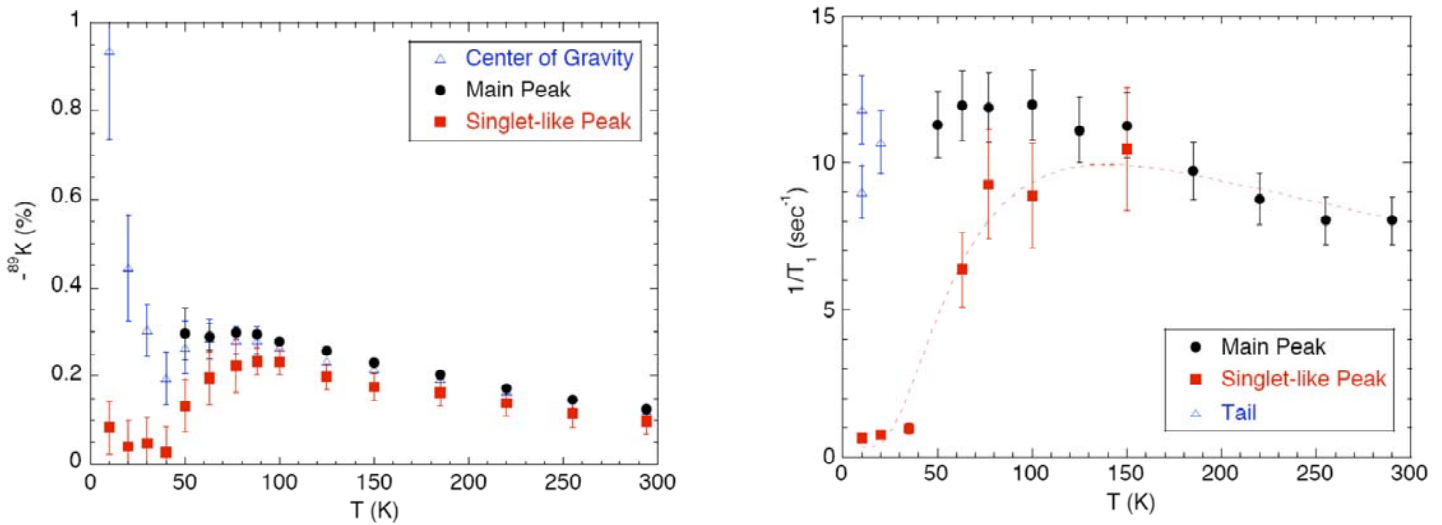

Figure 15. Temperature dependence of the paramagnetic Knight shift, $-^{89}$K,(Left) and the relaxation rate, $1/T_1$ (Right) for the "main" (lower frequency) peak and the singlet-like (higher frequency) peak of Fig.14(b). The dotted line is an empirical fit $1/T_1 \sim C/T \exp(-\Delta/k_B T)$ with $\Delta/k_B \sim 140$K. Integrated intensities of the two are roughly equal.(Color online)

In Fig.15,(left panel) the contrasting behavior of local magnetic susceptibility for the two distinctive $^{89}$Y environments in the sample is demonstrated by plotting the temperature dependence of $^{89}K$, plotted as -$^{89}K$ to reflect the fact that $A < 0$, defined for different components of the NMR lineshape. First, consider -$^{89}K_{CG}$, defined for the center of gravity of the whole lineshape (open triangles). In essence, -$^{89}K_{CG}$ is a bulk average of local magnetic susceptibility at the nuclear spin of each $^{89}$Y site, although the non-local nature of the hyperfine coupling makes the direct comparison somewhat non-trivial. Nonetheless, notice that the observed temperature dependence of -$^{89}K_{CG}$ is qualitatively similar to the SQUID data in Figure 10. -$^{89}K_{CG}$, monotonically increases from 295 K to ~100 K, levels off, then increases rapidly as T → 0. The Knight shift -$^{89}K_{Main}$ defined for the low frequency peak exhibits analogous behavior down to ~ 50 K, where the sharp feature becomes no longer observable in the NMR lineshape. In contrast, the local magnetic susceptibility at the location of $^{89}$Y nuclear spins involved in the higher frequency peak, as represented by -$^{89}K_{Singlet}$, begins to decrease below ~ 100 K down to ~ 40 K. The very small magnitude, -$^{89}K$ ~ 0.02 %, observed at 40 K suggests that the *local magnetic susceptibility is vanishingly small at these sites*. Recalling that the integrated intensity of the higher frequency peak is roughly half of the overall NMR lineshape, we conclude that about half of the magnetic moments at Mo sites become vanishingly small.

Further analysis of the -$^{89}K$ data can yield values of the hyperfine coupling constant, $A$, and an estimate of the separate contributions to the bulk susceptibility, $\chi = \chi_{spin} + \chi_{VV} + \chi_{dia}$, where $\chi_{VV}$ and $\chi_{dia}$ are the van Vleck and diamagnetic contributions, respectively. In Fig.16, -$^{89}K$ is plotted as a function of the bulk-averaged SQUID susceptibility, $\chi$, by choosing temperature as the implicit parameter. Note the linear relations between $-^{89}K$ and $\chi$ above ~ 100 K. From the slope, one obtains the hyperfine coupling constant $A = -23.4$ kOe/$\mu_B$ using the results of $-^{89}K_{CG}$. The magnitude of $A$ is ~ 5% greater (smaller) for $-^{89}K_{Main}$ ($-^{89}K_{Singlet}$). By extrapolating the linear fit to $-^{89}K_{CG} = 0$, we find that $\chi_{VV} + \chi_{dia} \sim 2.2 \times 10^{-4}$ emu/mol in the temperature range above ~ 100 K.

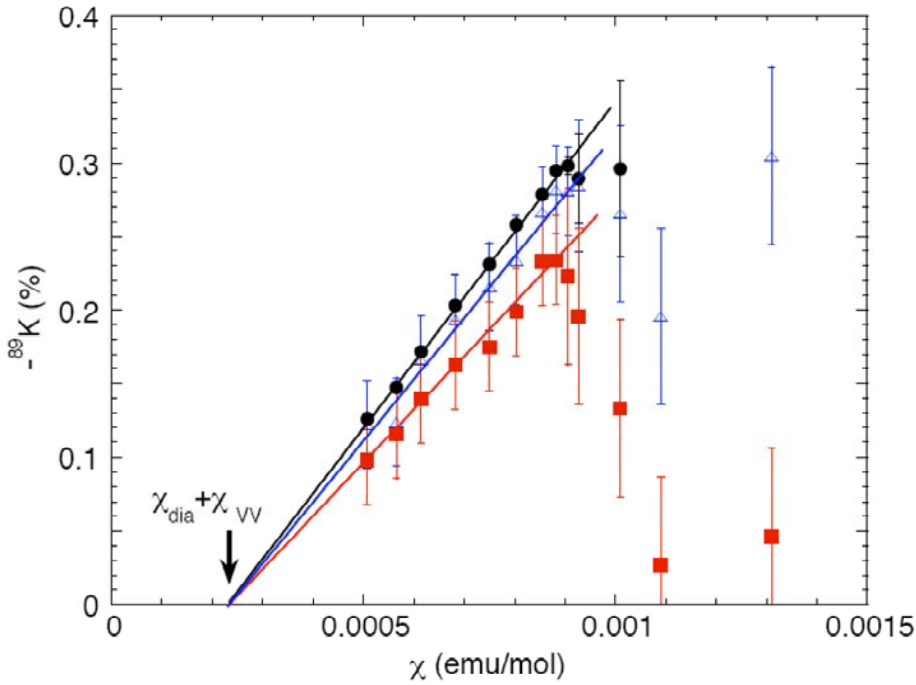

Figure 16. Scaling of the paramagnetic Knight shift ($-^{89}K$) with the bulk susceptibility for $Ba_2YMoO_6$. The derived hyperfine coupling constant, $A = -23.4 kOe/\mu_B$. The intercept gives the sum of the van Vleck and diamagnetic susceptibility components. (Color online)

Finally, the nuclear spin-lattice relaxation rate $1/T_1$ in Fig.15 (right panel) provides additional evidence for the presence of a collective singlet-like ground state of Mo magnetic moments. $1/T_1$ measures the spectral weight at the NMR frequency $^{89}f$ of the fluctuating hyperfine magnetic fields at the location of nuclear spins, and may be written

as $$\frac{1}{T_1} \propto T \sum_{\vec{q} \in 1st B.Z.} |A(\vec{q})|^2 \chi''(\vec{q}, f)$$

where $\chi''$ is the imaginary part of the dynamical susceptibility, **q** is the wave vector, f (~16 MHz) is the resonance frequency, $A(\mathbf{q})$ is the hyperfine form factor, and the summation over **q** should be taken over the first Brillouin zone..

If $1/T_1$ is dominated by fluctuating hyperfine magnetic fields from localized magnetic moments at temperatures much greater than the energy scales of interactions between themselves, we expect $1/T_1$ ~ constant [25]. On the other hand, if the magnetic moments develop short-range order, $1/T_1$ generally increases with decreasing temperature near a magnetic instability. These results for $1/T_1$ measured at the main sharp peak show only a mild increase from 295 K down to ~50 K. The absence of a pronounced peak or divergent behavior rules out the presence of any magnetic long-range order. However, $1/T_1$ measured at the singlet-like peak begins to deviate from the behavior of the main sharp peak below ~100 K, and rapidly tends toward zero. This implies that the low energy excitations of magnetic moments are nearly non-existent, as expected for a collective singlet ground state. In order to characterize this singlet like spin state, we model the overall temperature dependence as $1/T_1$ ~ $C/T \exp(-\Delta/k_B T)$, where C is a constant. We found that the behaviors of the singlet-like peak below 150K as well as the main peak above 150K can be reproduced well by choosing the gap size as $\Delta/k_B$ ~ 140K. Recall, however, that the singlet-like peak accounts for only about a half of all $^{89}$Y nuclear spins in the sample the other half being in the broad tail of the NMR lineshape below ~50 K, as shown in Fig.1(b). $1/T_1$ measured for these $^{89}$Y nuclear spins does *not* show any suppression below ~50 K, as shown by data points represented by open triangles. In other words, approximately half of $^{89}$Y nuclear spins continue to sense the same level of fluctuating hyperfine magnetic fields arising from low energy excitations of magnetic moments even below 50 K.

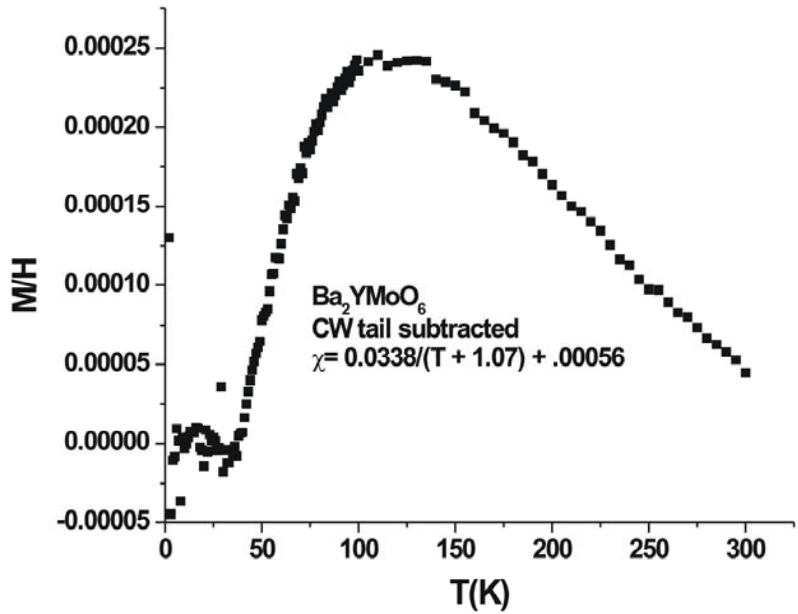

Figure 17. The Curie-tail subtracted bulk susceptibility of $Ba_2YMoO_6$.

It is of considerable interest to return to the bulk susceptibility at this stage, to look for evidence for the singlet state. In Figure 17, the bulk data are plotted following the subtraction of a Curie-Weiss plus TIP tail obtained by fitting the data below 40K. The constants for this fit are: C = 0.0338 emu-K/mol, θ = -1.07 K and χ(TIP) = 5.6 x $10^{-4}$ emu/mol. Note the similarity to the data of either Fig. 15 or Fig. 17, bringing the bulk and local susceptibility results into at least qualitative agreement. However, the Curie constant for the tail part of the bulk data, 0.0338 emu-K/mol, is much smaller than that found from fitting the high temperature bulk data, C = 0.29 emu-K/mol, and does not approach 50% as might be expected from the analysis of the local susceptibility via NMR.

Before continuing, it is important to consider a possible role for single ion physics in this system, as the $Mo^{5+}$ ion, $5d^1$, $t_{2g}^1$, has been shown to reside at a site of rigorously octahedral symmetry. This problem was first addressed by Kotani and and a detailed discussion can be found as well from other sources.[26,27] The basic result is that the perturbation of spin-orbit coupling on the $^2T_2$ crystal field term of the $t_{2g}^1$ configuration results in an unusual "non-magnetic quartet" (NMQ) ground state which arises due to an accidental cancellation of the spin moment by the unquenched orbital moment. This is not a singlet state in the sense that this term is normally used, but the result is that the ground state magnetic moment is in fact zero. This NMQ state is separated from a magnetic doublet by an energy gap $\Delta = 3\lambda/2$, where $\lambda$ is the single ion spin-orbit coupling constant. For $Mo^{5+}$, $\lambda = 1030$ cm$^{-1}$ or 1481 K, thus, a gap of $\Delta/k_B \sim 2200$ K would be expected on this basis. Even considering a reduction in $\lambda$ due to orbital delocalization effects of 20- 30 % or even more, the single ion energy scale is more than one order of magnitude larger than anything seen in the data presented above. Thus, while the single ion effects should be considered in any detailed theory for this material, it seems highly doubtful that the observed singlet state is of single ion origin and strengthens the case for a collective singlet state.

**Summary and Comparison with other S=3/2, S=1 and S = 1/2 systems.**

This is the third part of a systematic study of the ground states of geometrically frustrated B-site ordered double perovskites. Beginning with the monoclinic phases, the S=3/2 compound, $La_2LiRuO_6$ shows antiferromagnetic long range ordering[6], while the S=1 compound, $La_2LiReO_6$, finds a collective singlet ground state with a finite concentration of defects which can be polarized in applied fields, indicating a role for the reduced spin quantum number in ground state determination. [7] Remarkably, the S = ½ analog, $La_2LiMoO_6$, is nearly long range ordered, showing at least short range order in both the heat capacity and μSR behavior. This is somewhat surprising as, normally, one expects unconventional behavior for S = 1/2 due to the enhanced importance of quantum fluctuations. This observation is especially puzzling when compared to the isostructural, iso-spin material, $Sr_2CaReO_6$, which has a spin frozen ground state.[11]

A possible explanation of this apparent paradox is provided by consideration of the role of the local environment of the S = 1/2 magnetic ion in the two compounds. As already mentioned, the Mo – O coordination is octahedral with a weak tetragonal compression. In contrast the Re – O octahedron shows a weak tetragonal elongation. [11] Thus, in $La_2LiMoO_6$ one expects an isolated $d_{xy}$ ground state with nearly degenerate $d_{xz}$ and $d_{yz}$ at higher energies. Just the opposite level ordering will occur in $Sr_2CaReO_6$. This should give rise to a significant difference in the relative magnitudes of the various exchange pathways in the two materials. To test this hypothesis a spin dimer analysis was undertaken. [21,22] There are four such exchange pathways $J_1$ to $J_4$ as shown in Figure 19. Calculations were carried out assuming occupation of only the $d_{xy}$ orbital for $La_2LiMoO_6$ and equal occupation of $d_{xz}$ and $d_{yz}$ for $Sr_2CaReO_6$ and the results are displayed in Table 3. Note that for $La_2LiMoO_6$ the $J_4$ pathway exceeds the other three by at least an order of magnitude, while for $Sr_2CaReO_6$ there are three interactions of comparable magnitude, perhaps all four. Thus, these results suggest that a low dimensional model is more appropriate for the Mo phase while a geometrically frustrated model is better for the Re material. This is consistent with the observations, as, for a low dimensional magnet, strong, short range spin-spin correlations can develop which will be detected by a local probe such as µSR (Mo case), while geometric frustration is favored when the exchange pathways are of roughly equal strength (Re-case).

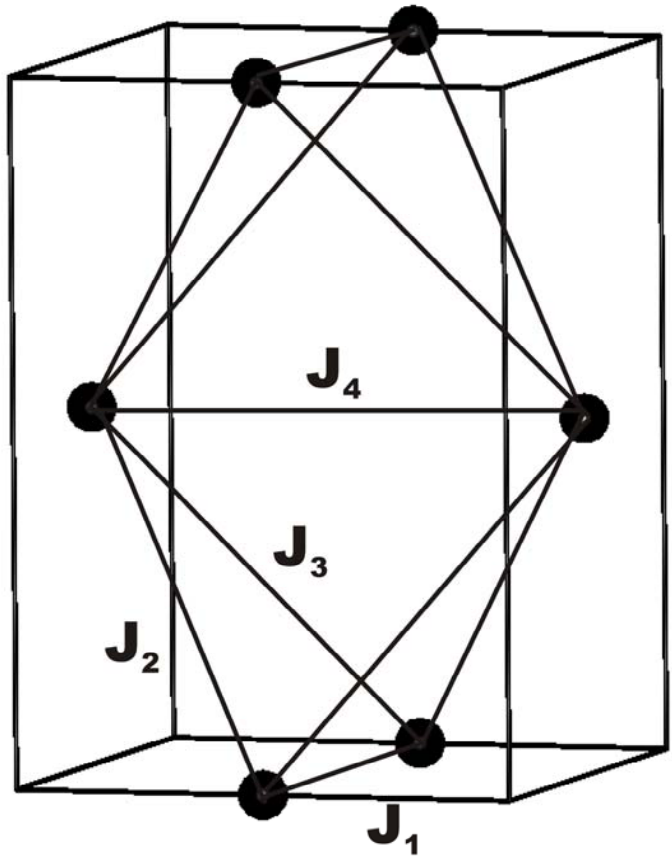

Figure 18. $Mo^{5+}(Re^{6+})$ sites in $La_2LiMoO_6$ ($Sr_2CaReO_6$) showing two edge-sharing tetrahedra within the monoclinic unit cell. The four exchange pathways calculated in Table 3 are indicated.

Table 3. Comparison of relative exchange pathway (Figure 18) strengths in $La_2LiMoO_6$ and $Sr_2CaReO_6$.

| Pathway | $La_2LiMoO_6$ ($d_{xy}$) | $Sr_2CaReO_6$ ($d_{xz}, d_{yz}$) |
|---|---|---|
| $J_1$ | 0.14 | 1.0 |
| $J_2$ | 0.014 | 0.16 |
| $J_3$ | $4.3 \times 10^{-4}$ | 0.25 |
| $J_4$ | 1.0 | 0.87 |

Turning to the cubic materials, $Ba_2YRuO_6$ (S = 3/2) shows a large frustration index, f ~ 18, but, in spite of detectable Y/Ru site mixing at the 1% level, is long range AF ordered below 36K.[6] Cubic $Ba_2YReO_6$(S = 1) is also highly frustrated, f > 12, with no detectable Y/Re site disorder, yet, it shows spin freezing behavior with an ill defined freezing temperature. On the other hand, cubic $Ba_2YMoO_6$ (S = ½) exhibits a most unusual heterogeneous ground state comprised, roughly equally, of gapped, collective spin singlet and paramagnetic components. In spite of a ~ 3% Y/Mo site mixing there is no sign of spin freezing on any time scale investigated to 2K. A singlet/triplet gap, $\Delta/k_B$ ~ 140K was inferred from NMR data. Neutron inelastic scattering experiments are currently underway to investigate further the nature of the gap. Among these three iso-structural compounds, the enhanced quantum fluctuations in the S = 1 and S = ½ materials appear to play an important role in suppression of long range order.

Comparing $Ba_2YMoO_6$ to other S = ½ cubic double perovskites, one notes immediately a sharp contrast with $Ba_2LiOsO_6$ and $Ba_2NaOsO_6$ (based on $Os^{7+}$) which are antiferromagnetic and ferromagnetic, respectively, with minimal evidence for a role for geometric frustration. [12] This is a most puzzling situation as no clear pattern of behavior emerges for these iso-structural/iso-spin materials which indicates that a detailed understanding of these remarkable differences will not be easily gained.

**Acknowledgements.**

We thank Paul Dube for assistance with SQUID and heat capacity measurements and H.F. Gibbs for TGA data. Franco Ieropoli assisted with the early stages of this work. J.E.G, G.M.L. B.D.G. and S.K. thank NSERC of Canada for support of this research and V.K.M. acknowledges a NSERC postgraduate fellowship. S.K. thanks the Canadian Foundation for Innovation and B.D.G., G.M.L. and T.I. thank the Canadian Institute for Advanced Research. For work done at Florida State University, C.R.W. thanks the NSF for funding through grant number DMR-08-04173.